\def\bey{\begin{eqnarray}}
\def\eey{\end{eqnarray}}
\def\be{\begin{equation}}
\def\ee{\end{equation}}
\def\ba{\begin{array}}
\def\ea{\end{array}}

\def\ld{\lambda}
\def\Ld{\Lambda}
\def\af{\alpha}
\def\sg{\sigma}

\def\om{\omega}

\def\bt{\beta}

\def\nnb{\nonumber}

\bigskip
\documentclass[twocolumn,preprintnumbers,amsmath,amssymb]{revtex4}
\usepackage{graphicx}
\usepackage{fancyhdr}
\usepackage{dcolumn}
\usepackage{bm}
\usepackage{amssymb}
\usepackage{amsmath}
\usepackage{graphicx}
\usepackage{float}
\usepackage[normalem]{ulem}
\usepackage[dvips]{color}
\usepackage[colorlinks=true,citecolor=blue,linkcolor=black]{hyperref}

\UseRawInputEncoding
\begin{document}
\preprint{ }
\title{Neutron star deformability with hyperonization in
density-dependent relativistic mean-field models }
\author{\quad W. Z. Shangguan$^{1,2}$\footnote{%
Visiting scholar at the first affiliation since the year of
2020.},
Z. Q. Huang$^{2}$, S. N. Wei$^{1}$, W. Z. Jiang$^{1}$\footnote{%
Corresponding author: wzjiang@seu.edu.cn}}
\affiliation{$^1$ School of Physics, Southeast University, Nanjing 211189, China\\
$^2$ Faculty of Pharmacy, GuangXi University of Chinese Medicine,
Nanning 530200, China }

\begin{abstract}
Neutron star tidal deformability extracted from
gravitational wave data provides a novel probe to the interior
neutron star structures and the associated nuclear equation of
state (EOS). Instead of the popular composition of nucleons and
leptons in neutron stars, we include hyperons and examine the role
of hyperons in the tidal deformability and its impact on the
symmetry energy in a relativistic mean-field approach with the
density-dependent parametrizations.  The hyperons are found to
have significant impact on the deformability, correlated
sensitively with the onset density and fraction of hyperons in
neutron star matter. Moderately lower onset density of hyperons
can yield considerable modification to the tidal deformability
and shift its inference on the nuclear EOS.    The future measurements of the tidal deformability at multi-fiducial star masses are anticipated to lift the degeneracy between  the contributions from the hyperon component and symmetry energy. 
\end{abstract}


\maketitle

\section{Introduction}
According to the general relativity, the moving body changes the
surrounding geometry, and the reciprocal motion or transient
mass-change process can cause the sonic propagation of space-time
oscillation, the so-called gravitational wave. The massive
celestial  binary merger  is one of strongest
sources of gravitational waves. It was fascinating that the
gravitational wave signal from the GW170817, which is a neutron star
merger with a chirp mass 1.188  $M_\odot$ and 40 Mpc away from the
Earth,  was successfully detected by Ligo and Virgo detectors~\cite{Abb17,Abb18}   more than 3 years ago.  It is a historic event in the deep-sky detection. Its significance also exists in nuclear
physics. Since then, the study on neutron stars heads into a
multi-messenger era. The tidal deformability measures the size of
neutron stars and is correlated with the neutron distribution of
heavy nuclei~\cite{Far18,De18}.  In particular,    during the inspiral of the binaries close to merger, 
the tidal deformability of neutron stars encodes the information of the equation of state (EOS) of asymmetric matter in the interior neutron star.

The tidal deformability (or, alternatively the Love
number) of neutron stars can be derived as a metric
perturbation in the general relativity~\cite{Tho67,Hin08,Dam09}, since  matter is the source of the metric. It was found that the
Love number of normal neutron stars is quite different from that
of strange quark matter stars~\cite{Hin10,Pos10}.  Several research groups have engaged in exploring the tidal deformability with its
possible constraint on  the EOS of asymmetric matter. The
uncertainty of the symmetry energy at saturation density just has
a moderate effect on the tidal deformability~\cite{Far13}.
Recently, the tidal deformability from the experiments is
optimistically used to constrain the symmetry energy at
suprasaturation densities~\cite{Zha18,Ton20}.

Among all non-nucleonic degrees of freedom in neutron stars, the
hyperon is an important but contentious ingredient. Usually, the
inclusion of hyperons can clearly soften the nuclear EOS and
reduce the maximum mass of neutron stars significantly, for
instance, see Refs.~\cite{Jia12,Wei19}. It was even claimed that
the observations of large-mass neutron stars (for instance,
pulsars J1614-2230~\cite{Dem10} and J0740+6620~\cite{Cro20}) seem
to rule out the hyperon EOS. On the other hand, quite a few
hyperon EOS's were proposed to reproduce the 2$M_\odot$ neutron
stars~\cite{Kr2009,Be2011,Ta02,Ts2009,Ne18,Fo20}. One of the
authors and his collaborators also worked out the hyperon EOS that
is compatible with the mass and meantime the radius constraints of
neutron stars~\cite{Jia12}  and is in the rank of fine models, see Ref.~\cite{Cha16} and references therein. 
In spite of the dispute of the
existence problem,  the onset densities of hyperons and their
fractions also diversify rather largely in a variety of models.
These issues are rooted in the in-medium interactions for hyperons
and remain largely unsolved.  In a multi-messenger era, the
gravitational wave signals would  hopefully light the secrets for
the hyperon component in stars and the underlying interactions,
although no much attention has been paid to the accordance between
the available hyperon EOS's and the tidal deformability of the
1.4$M_\odot$ stars  for simplicity or due to the absence of the hyperon involvement in extracting the EOS of asymmetric matter from terrestrial experiments. However, the seemly small chirp mass   (1.188 $M_\odot$) of the GW170817 with the component masses ranging from 1.17 to 1.6 $M_\odot$ does not ever mean  that the hyperon component is negligible.
In this work, we will aim to scrutinize the interplay between the tidal deformability and the hyperon EOS
based on the previous density dependent relativistic mean-field
(RMF) models~\cite{Jia07,Jia07b}. In order to single out the
hyperon effect beyond saturation density, we need first pin down
the EOS in the low density region where there is a transition from
the interior homogeneous phase to inhomogeneous crustal phase.  We
will determine the transition density by   the
instability condition of uniform matter~\cite{Lat07,Wei17}  and further distinguish the inner and
outer crusts by appropriately choosing different EOS's for the
inner and outer crusts~\cite{Ca03}.  With moderately tuned onset
densities and fractions of hyperons, we can observe the
significant role of hyperons in affecting the tidal deformability of intermediate-mass neutron stars  
and the extraction of nuclear symmetry energy beyond saturation
density.

The remaining of the paper is organised as follows. In the subsequent
section, a brief formalism is presented for the differential
equation of the Love number integrated in the Tolman-
Oppenheimer-Volkoff (TOV) equation and the RMF EOS. The emphasis
is placed on the corresponding parametrization of nuclear EOS
concerning the hyperon interactions. In sec. III, we present
numerical results and analyze the hyperon effect on the tidal
deformability and  the nuclear EOS. Finally, a brief summary is given in Sec. IV.

\section{Formalism and parametrizations}

The quadrupole tidal field can be  incorporated in the spacetime
metric   as an external perturbation specified by a function $H$
which satisfies the following differential equation~\cite{Hin08}:
\bey
 H^{\prime\prime}(r)&+&\left[\frac{2}{r}+e^{\ld(r)}\left(
 \frac{2M(r)}{r^2}+4\pi r (p(r)-\mathcal{E}(r)) \right)
 \right]\times  \nnb \\
 && H^{\prime}(r)+H(r)Q(r)=0,\label{eqH}
 \eey
where \bey
 Q(r)&=&4\pi e^{\ld(r)}\left( 5\mathcal{E}(r)+9p(r)+\frac{d\mathcal{E}(r)}{dp(r)}
  (p(r)+ \mathcal{E}(r))
 \right) \nnb \\
  &&-6\frac{e^{\ld(r)}}{r^2}-(\nu^\prime(r))^2,
  \eey
and the metric functions $\ld(r)$ and $\nu(r)$ are given as
 \be e^{\ld(r)}=\left[1-\frac{2M(r)}{r}\right]^{-1},\hbox{ }
     \nu^\prime(r)=2e^{\ld(r)}\frac{M(r)+4\pi r^3 p(r)}{r^2},
      \ee
with $M(r)$, $p(r)$, and $\mathcal{E}(r)$ being the mass,
pressure, and energy density, respectively. By redefining the
quantity $y=H^\prime/H$, Eq.(\ref{eqH}) turns out to be following
first-order differential equation:
 \be
 y^\prime(r)+y^2(r)+F(r)y(r)+Q(r)r^2=0,\label{eqy1}
 \ee
where
 \be
 F(r)=e^{\ld(r)}[1+4\pi r^2(p(r)-\mathcal{E}(r))],
  \ee
  with $ y(0)=2.$
The Love number $k_2$ is obtained at neutron star surface with
$y_R=y(R)$, and it is given by
 \bey
 k_2(y_R)&=&\frac{8}{5} \bt^5 (1-2\bt)^2 [2-y_R+2\bt(y_R-1)]\{2\bt
  \nnb \\
  & &[6-3y_R+3\bt(5y_R-8)]+4\bt^3[13-11y_R \nnb\\
  & &+\bt(3y_R-2)+  2\bt^2(1+y_R)]+3(1-2\bt)^2\nnb\\
  & &[2-y_R+2\bt(y_R-1)]\ln(1-2\bt) \}^{-1}, \label{eqk2}
  \eey
where $\bt=M/R$ is the dimensionless compactness parameter in unit
of $G=c=1$. The tidal deformability is given by
 \be\label{eqldg}
 \Ld_g= \frac{2}{3}k_2\left(\frac{R}{M}\right)^5,
  \ee
  with $R$ and $M$ being the radius and
mass of the neutron star, respectively.

Equation (\ref{eqy1}) for the perturbation tidal field should be solved
together with the TOV equations:
 \bey
p^\prime(r)=-\nu^\prime[p(r)+\mathcal{E}(r)]/2,\hbox{
}M^\prime(r)=4\pi r^2\mathcal{E}(r),
 \eey
which are solved by integrating over radial coordinate from the
star center to the surface where the pressure vanishes. We
perform the integration with the fourth-order Runge-Kutta method.
The nuclear EOS, i.e., $p(\mathcal{E})$ is an input of the
integration.  The central energy density or pressure is chosen as
a free parameter to obtain a mass-radius trajectory for neutron
stars.

In obtaining the deformability, conveniently and popularly
used method relies on the EOS with a simple neutron star composition of nucleons and leptons or a polytropic piece-wise EOS, whereas we
start the work from a Lagrangian that consists of the fields of
baryons, leptons ($e,\mu$), and mesons, and the
interactions between them. Here, we invoke directly the energy
density and pressure from the previous density-dependent  RMF models~\cite{Jia12}:
\begin{eqnarray}
{\mathcal{E}}&=&\frac{1}{2} m_\omega^{*2} \omega_0^2 +\frac{1}{2}
m_\rho^{*2} b_{0}^2 +\frac{1}{2} m_\phi^{2} \phi_0^2+
 \frac{1}{2} m_\sigma^{*2}\sigma^2 \nonumber\\
 && + \frac{1}{2} m_{\sigma^*}^{2}{\sigma^*}^2 +2\sum_{i}
\int_{0}^{{k_F}_i}\! \frac{d^3\!k}{(2\pi)^3 }~ E_i^*,  \label{eqe1} \\
 p&=&\frac{1}{2} m_\omega^{*2} \omega_0^2
+\frac{1}{2} m_\rho^{*2} b_{0}^2+ \frac{1}{2} m_\phi^{2} \phi_0^2
-\frac{1}{2} m_\sigma^{*2}\sigma^2 -\Sigma^R_0\rho \nonumber\\
&&  -\frac{1}{2} m_{\sigma^*}^{2}{\sigma^*}^2
  +  \frac{2}{3}\sum_{i}\int_{0}^{{k_F}_i}\!\! \frac{d^3\!k}{(2\pi)^3}
         \frac{{\bf k}^2}{E^*_i},
\label{eqp1}
\end{eqnarray}
where $i$ runs over the species of baryons and leptons considered
in neutron star matter, $E^*_i=\sqrt{{\bf k}^2+m_i^{*2}}$ with
$m_i^*$ being the Fermion effective mass, and
$\Sigma^{R}_0$ is the rearrangement term, originated from the
density-dependent parameters. The explicit formula of the
rearrangement term can be referred to Ref.~\cite{Jia13}. The meson
coupling constants and masses with asterisks denote the density
dependence, given by the Brown-Rho (BR)
scaling functions~\cite{Jia07,Jia07b,Jia12}. It is interesting to note that the parametrization with this density dependence  respects the chiral limit in terms of the vanishing scalar density and nucleon effective mass at high densities,
 which is interpreted as the vector manifestation of chiral symmetry in the hidden local symmetry theory~\cite{GE04ab}.  
In the present work, the RMF  parameter sets SLC and
SLCd~\cite{Jia12,Jia07b} that can  reproduce  the ground-state properties of finite nuclei and meet  the 2$M_\odot$ constraint of neutron stars are adopted to study neutron stars with hyperonization, and  the composition of neutron stars  consists of  baryons $(N,\Lambda,\Sigma, \Xi)$ and leptons
$(e,\mu)$.

For the hyperonic sector, the strange mesons $\phi$  (1020MeV) and $\sigma^*$ (i.e. $f_0$, 975MeV), in addition to normal mesons, are included  with their  parameters free of density~\cite{Jia12}.
The coupling of hyperons with normal mesons can generally be specified by the ratios of the meson coupling with hyperons
to that with nucleons: $X_{i Y}=g_{iY}/g_{iN}$ with $i$ denoting meson species. Although  these  ratio parameters are, in most cases, taken to be constants in the literature,  they are being  density-dependent ones $X_{iY}(\rho)$ for the  scaling functions for hyperons~\cite{Jia12}:
\begin{eqnarray}\label{eqphiY}
&&\Phi_{\omega \Lambda(\Sigma)}(\rho)=(\frac{1}{3}-\af)\Phi_{\omega
N}(\rho_0)
+(\frac{2}{3}+\af)\Phi_{\omega N}(\rho),\nonumber\\
&&\Phi_{\omega \Xi}(\rho)=(\frac{2}{3}-\af)\Phi_{\omega N}(\rho_0)
+(\frac{1}{3}+\af)\Phi_{\omega N}(\rho),\\
&&\Phi_{\sigma Y}(\rho)=(1-f_{\sigma Y})\Phi_{\sigma N}(\rho_0)
+f_{\sigma Y}\Phi_{\sigma N}(\rho),\nonumber
\end{eqnarray}
where $\Phi_{iN}(\rho)$ are the nucleon scaling functions, $\rho_0$ is the saturation density (0.16fm$^{-3}$), and
$f_{\sigma Y}$ and $\alpha$ are adjustable constants. The scaling function
$\Phi_{\rho\Xi}$ for the $\rho$ meson takes the same as that
of the $\omega$ meson. The  product of the free-space meson-baryon coupling constant and the scaling function defines the coupling constant at each density. In Eq.(\ref{eqphiY}), the parameter
$\af$ is newly invoked to tune the density dependence in the vector meson couplings, which is relevant to the in-medium effect from the hyperonic sector. This small parameter  can be
used to adjust the onset density and fractions of hyperons in
neutron star matter efficiently.  The two free parameters $\af$ and $f_{\sg Y}$  do not change the hyperon potentials at saturation
density that are set as the empirical
values~\citep{Jia12,Ha1989,Fu1998}
\begin{equation}\label{eqhpot}
U^{(N)}_\Lambda=-30 MeV=-U^{(N)}_\Sigma, \hbox{ } U^{(N)}_\Xi=-18 MeV.
\end{equation}
The free-space parameters concerning the hyperons
are the same as those in Table 1 of Ref.~\cite{Jia12},
regardless of the new parameter $\af$.
Note that the choice of $U^{(N)}_\Sigma(\rho_0)$ has some arbitrariness for uncertainty~\cite{Fo20},  
and in the present models
the $\Sigma$ hyperons actually do not appear for any repulsive potential. Similar expulsion of $\Sigma$ hyperons was also revealed in the RMF model GM1~\cite{Sch08}.

In the low density region, there are no hyperons and even no
muons. The EOS  of this density region comprises of two pieces:
the inner and outer crustal ones. In the inner crust, we adopt a
phenomenological EOS
$p(r)=a+b\mathcal{E}(r)^{4/3}$~\cite{Link99,Ca03} with constants
$a$ and $b$ being determined by the continuous condition at the
core-crust transition density and the density
$\rho=2.57\times10^{-4}$ fm$^{-3}$ with the energy density
$\mathcal{E}=0.24$ MeV fm$^{-3}$ and $p=4.87\times10^{-4}$  MeV
fm$^{-3}$ which is a point connecting  to the outer crust~\cite{Ca03,Bay71}. The
core-crust transition density $\rho_t$ is here determined as the lowest density of uniform phase by the
stability of matter that requires the convex energy  against the volume~\cite{Lat07,Wei17},  and it is 0.0912 and 0.0928 fm$^{-3}$ for SLC and SLCd, respectively.  In addition to the above continuous  connection, high-order discontinuities may still exist at the core-crust interface. Here, we follow the method in Ref.~\cite{Pos10} to deal with the  discontinuity in $d\mathcal{E}/dp$ using the Dirac delta function. We have noticed the work by Piekarewicz
et al that a continuous first derivative of pressure is imposed on
both interfaces of the inner crust~\cite{Pie19}.  For the outer
crust, we employ the empirical EOS given by Baym et
al~\cite{Bay71}.  Eventually, we adopt the piece-wise EOS's for neutron stars with the interfacial matching specified above.  It is worth mentioning that Fortin et al studied systematically  the uncertainty in the crust thickness and star radius arising from a variety of core-crust EOS matchings~\cite{Fo16}. Similar study combined with the GW170817 data was later performed by Ji et al~\cite{Ji19}. In this work, we use the same matching scheme to focus on the hyperon contribution concerning the core EOS. For comparison, we also examine
the case with the total crustal EOS of Ref.~\cite{Bay71} below $\rho_t$,   but find a negligible deviation from the one herein.

\section{Numerical results and discussions}

The discovery of large-mass neutron stars imposed the challenge to
the hyperon EOS for neutron stars. In the previous work, the
hyperon EOS survives in the large-mass neutron stars by invoking
the density-dependent nucleon-hyperon interactions which allow the
hyperons to reside in a shell in the interior of neutron
stars~\cite{Jia12}. In a multi-messenger era, it is necessary to
check whether such a hyperon EOS is compatible with the neutron
star tidal deformability extracted from data years ago.

Prior to the discussion of the numerical results, we first interpret the RMF models SLC and SLCd briefly. These two models can reproduce the ground-state properties of finite nuclei fairly well. The only
difference of the two models is that the SLCd has a softer
symmetry energy at high densities than that of the SLC. The slope
parameter $L$ of the symmetry energy at saturation density is
$92.3$ and $61.5$ MeV for the SLC and SLCd, respectively.
These values are within or close to  some globally averaged values $59\pm 16$ MeV~\cite{Li13}. 
For the slope parameter $L$, there are also clearly lower  ranges either extracted from data~\cite{Lat13,Lat14} or obtained from the ab initio results of neutron matter~\cite{Heb13}. With inclusion of the clearly lower $L$ range, an average of the $L$ values gives  a larger range of $58.7\pm 28.1$ MeV~\cite{Oe16}.
Note that the latest measurement of $^{208}$Pb neutron skin thickness ($0.283\pm0.071$ fm) through the weak-interaction probe~\cite{PREX}   would suggest significantly larger value of  $L=106\pm37$ MeV~\cite{Re21}. Very recently, a large span of the $L$ with an upper bound 117 MeV was extracted from the spectra of charged pions~\cite{Es21}.
In these cases, the value of $L=61.5$ MeV with the SLCd would be near the lower bound, and the value $92.3$ MeV with the SLC should be still well within the experimental bounds.   The neutron skins of $^{208}$Pb are 0.21
and 0.17 fm with SLC and SLCd, respectively~\cite{Jia07b}, which
agree satisfactorily with values extracted from various
experiments. It was found that the symmetry energy is associated
with the onset density of hyperons~\cite{Jia06}. The soft symmetry
energy leads to the smaller neutron chemical potential and
consequently the smaller chemical potential of $\Lambda $ hyperon
in chemical equilibrium. The threshold density for the hyperon
onset has to be larger so that the required minimum neutron
chemical potential can be reached. Thus, we will find that the
onset density of hyperons with the SLCd is larger than that with
the SLC.
\begin{table}[tbh]
\caption{Onset densities of the $\Lambda $ hyperon for various choices of $
\protect\alpha $ and $f_{\protect\sigma \Lambda }$ with the SLC
and SLCd. The density is in unit of $\protect\rho _{0}$ that is
0.16 fm$^{-3}$. } \label{tab1}
\begin{center}
\begin{tabular}{ccccc}
\hline\hline
Model & $f_{\sigma\Lambda}$ & $\alpha=0$ & $\alpha=-0.05$ & $\alpha=0.05$ \\
\hline
SLC & 0.8 & 2.63 & 2.70 & 2.57 \\
& 0.9 & 2.78 & 2.88 & 2.69 \\ \hline
SLCd & 0.8 & 2.85 & 2.97 & 2.77 \\
& 0.9 & 3.06 & 3.24 & 2.93 \\ \hline\hline
\end{tabular}%
\end{center}
\end{table}

In addition to the symmetry energy, the parameter $\alpha $ in Eq.(\ref{eqphiY}) can also shift the hyperon onset density. The negative value of $\alpha $ ramps up the fraction of the density-independent part of
$\Phi _{\omega Y}$, and increases the hyperon chemical potential
and consequently the onset density. The positive value of $\alpha
$ shifts them on the opposite. In Table~\ref{tab1}, we present the
onset densities of $\Lambda $ hyperons in different cases. In
fact, the hyperon onset density can't be detected directly and is
very different with various interactions in different models,
ranging from about $2\rho _{0}$ to $4\rho _{0}$. For instance, In
usual RMF models, the hyperon onset density locates roughly at
twice normal density~\cite{Jia06}. In the nonlinear
self-interactions involving a vector meson with hidden
strangeness, the onset density of hyperons can be as high as
$3\rho _{0}$ arising with a suppressed hyperon
fraction~\cite{Be2011}. Such a suppression with larger onset
densities can also be given by invoking a new boson coupling with
hyperons~\cite{Kr2009}. In Refs.~\cite{Ta02,Ts2009}, hyperons were
found to arise above $4\rho _{0}$ with a rather limited effect on
the EOS of neutron star matter. As one can see in
Table~\ref{tab1}, the onset densities in our work are above
2.5 $\rho _{0}$, and the small tuning of the parameter $\alpha $
away from naught yields the moderate shift in the onset density
within 0.3$\rho _{0}$ that is just moderate, compared to the large
diversity presented in the literature.

With the appearance of the hyperons, the EOS becomes softened as
naturally given by the stability of matter. This usually results in
the considerable reduction of the maximum mass of neutron stars.
For instance, with the constant ratio parameters
$X_{i Y}$  in the present RMF models, the maximum
mass of neutron stars is just as high as
1.4$M_\odot$~\cite{Jia12}, which is obviously against the
observation of large-mass neutron stars. However, it was
revealed~\cite{Jia12} that the softening can evolve in density
consecutively to a stiffening  by invoking the density-dependent
hyperonic interaction in terms of the density dependent ratio
parameters $X_{i Y}$. More specifically, the occurrence of the stiffening results dominantly from the repulsion provided by the $\omega$ meson through the ratio parameter $X_{\om Y}$.
Note that with constant ratio parameters, the problem of the negative nucleon effective mass, encountered at high densities in hyperonized matter, has to be treated by necessarily connecting to the quark matter EOS  prior to the occurrence.  There is no such problem for the density dependent ratio parameters that are adopted for hyperons in this work.  
While the stiffening comes up with the suppression of hyperon fractions, the neutron star matter can transit for stability to the normal isospin-asymmetric matter prior to the vanishing of
hyperons.

\begin{figure}[thb]
\begin{center}
\includegraphics[width=6.5cm]{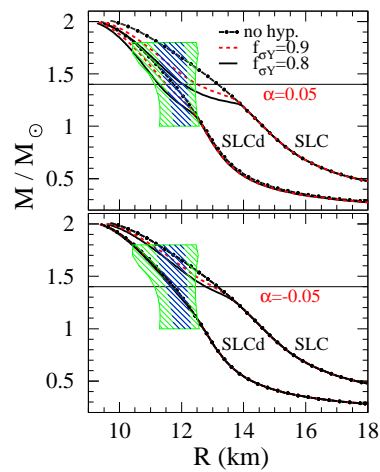}
\caption{\label{fmsr} (Color online) The mass-radius relation of
neutron stars. The parametrizations for various curves are
either labelled explicitly or specified in Table~\ref{tab1}. The
star composition of the case without hyperons includes the
nucleons, electrons and muons. The hatched areas give the
probability distributions with $1\sigma$ (blue) and $2\sigma$
(green) confidence limits~\cite{St2010}.}
\end{center}
\end{figure}
\begin{figure}[tbh]
\begin{center}
\includegraphics[width=7cm]{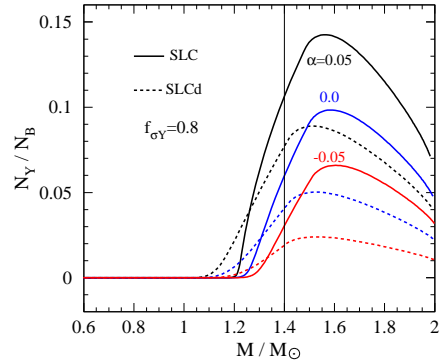}
\end{center}
\caption{(Color online) The hyperon number fraction as a function
of star mass. $N_Y$ and $N_B$ are the total hyperon and baryon numbers in the star, respectively. The curves are presented for two models SLC and SLCd with $f_{\sigma Y}=0.8$ and various $\af$ as labelled.}
\label{fhyp}
\end{figure}

With the EOS's specified, we can carry out the mass-radius
relation and tidal deformability parameters of neutron stars. In
Fig.~\ref{fmsr}, the mass-radius trajectories with the SLC and
SLCd are plotted for $\af=0.05$ and -0.05. The results with
$\af=0$ lie between the two cases and were given in Ref.~\cite{Jia12}.
We take the curve without hyperons as the fiducial case  and
measure the relative variation of the trajectories with various
parametrizations. As shown in Fig.~\ref{fmsr}, the parameter $\af$
plays a sensitive role in shifting the mass-radius trajectory and
radius separation between the normal neutron stars and hypernized
neutron stars. Such an $\af$-induced separation is related to the
various hyperon onset densities shifted by $\af$. Meanwhile, the
parameter $\af$ also induces the suppression or enhancement of
hyperon fractions corresponding to the larger or smaller hyperon
onset densities, respectively. As an evidence, we plot in
Fig.~\ref{fhyp} the total hyperon number fraction as a function of
the star mass. We see that the hyperon fraction, which is a
ratio of the  $\Ld$ plus $\Xi$ number over the total baryon
number, is generally small.  On the other hand, the curves in
Fig.~\ref{fhyp} have two more distinct features.
The first one is that the significant difference appears in the curves of the SLC and SLCd parametrizations with different peak positions.  
The smaller fraction with the SLCd is associated with its larger onset density, in contrast to that with the SLC.
The peaks in the curves  around 1.5$M_\odot$ with the SLCd and 1.55$M_\odot$ with the SLC arise as the  balance between the heavier star with more hyperons included and the  exclusion of the hyperon component in the high density region for the stiffening of the EOS, as mentioned above and referred to Ref.~\cite{Jia12} for more details. For instance,  an exclusion zone of the hyperons in the 1.5$M_\odot$ star with all the SLCd parametrizations is a sphere with a radius of  about 3 km from the star center, and the hyperon zone extends down to the low density region for about 4 km. For a 1.5$M_\odot$ star with the SLC, the exclusion zone in about a 1km radius around the center forms only for the parametrization with $\alpha=0.05$.   
Secondly, large difference also arises from different $\af$'s. With
more suppression induced by the parameter $\af$ further, the
neutron star radius with the SLCd runs almost out of the zone that
is sensitive to the hyperon composition, as shown in the lower
panel of Fig.~\ref{fmsr}.  As a result,  the  difference in star radii with the SLC and SLCd  reduces clearly by including the  hyperon fraction in neutron stars. This clear reduction is eventually  attributed to the only difference in two models, namely,  the density dependence of the symmetry energy. The softer symmetry energy in the SLCd increases the hyperon onset density and reduces the hyperon fraction significantly, while the SLC with a stiffer symmetry energy gives  a smaller  onset density with a clearly larger hyperon fraction. The inclusion of hyperons in SLC thus lowers  the pressure of matter significantly and results in a clear shrinkage of the star radius, which is consistent with the pressure-radius correlation  at intermediate densities ($1.5\rho_0<\rho<2-3\rho_0$)~\cite{Lat07}. Accordingly,  an appreciable reduction of the radius difference from two models is observed in Fig.~\ref{fmsr}.

\begin{figure}[thb]
\begin{center}
\includegraphics[width=9cm]{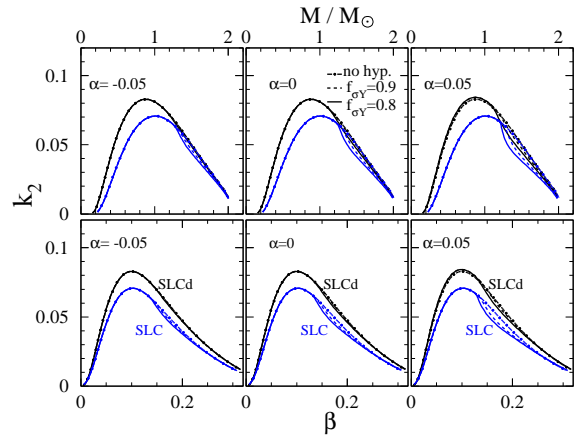}
\end{center}
\caption{(Color online) The Love number as a function of  star mass in unit of solar mass (upper panels) and compactness $\protect\beta=M/R$ (lower panels)
for $\af=-0.05,0, 0.05$. In each panel, the curves subject to the
models SLC and SLCd are respectively  presented for three cases:
without hyperon, $f_{\sigma Y}=0.8$ and 0.9.  } \label{fk2}
\end{figure}

\begin{figure}[thb]
\begin{center}
\includegraphics[width=6.5cm]{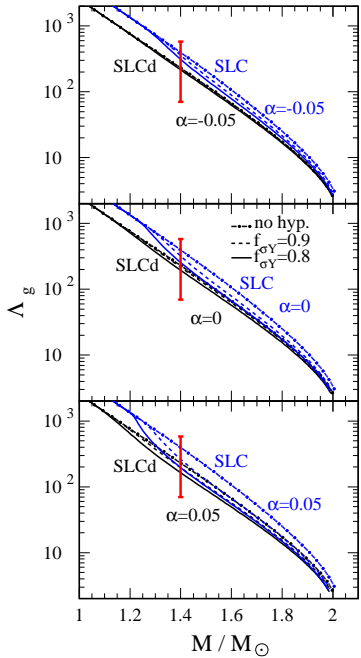}
\end{center}
\caption{(Color online) The tidal deformability as a function of
star mass with the same legend  as in Fig.~\ref{fk2}. The error
bar in red is the experimental bounds extracted for the
1.4$M_\odot$ neutron star~\protect\cite{Abb18}. } \label{fld3}
\end{figure}

The Love number $k_{2}$, which is carried out together with the
mass-radius relation  in a set of coupled equations,  
is shown in Fig.~\ref{fk2}. The
maximum value of $k_{2}$ is situated around 1$M_{\odot }$ for the SLC and 0.9$M_{\odot }$ for the SLCd, and  the difference in   $k_{2}$  between the two models
reaches the maximum around the peak regions. It is found that  the difference in  $k_2$ at the given star mass is correlated predominantly with the difference in the star central pressure (or, the central energy density), since the central pressure at the origin serves as the starting point with the largest resistance against the gravity and affects the density profile in the neutron star by integrating the TOV equations and  Eq.(\ref{eqy1}) outwards. For instance, the relative difference in the central energy densities of the two models decreases from about 24\%  for a $0.91M_\odot$ neutron star to about 4\% for a $1.3 M_\odot$ one. As shown in the upper panels in Fig.~\ref{fk2}, the difference in the star radii that can be specified by the one in the symmetry energies  between the two models just has very limited effect on the difference in $k_{2}$, especially, for neutron stars with $M>1.3M_\odot$, which is consistent with the result in Ref.~\cite{Far13}. Shown in the lower panels of Fig.~\ref{fk2} is the $k_2$ versus the compactness parameter $\beta$, and the difference in  $k_2$  from the two models at given $\beta$ can be specified by the different $y_R$'s in Eq.(\ref{eqk2}) that depend on the radius and central energy density both.   It is interesting to see that
the hyperon composition, albeit with small fractions in neutron
stars, can  shift the Love number to
some extent, especially in the case of the larger hyperon
fraction. This is shown rather clearly in the upper right panel
of Fig.~\ref{fk2} with $\alpha =0.05$ where the hyperon fraction
is at the top of three cases as shown in Fig.~\ref{fhyp}. Here, the relative shift of the Love number between two models arises from a moderate enhancement of the difference in central energy densities and the distinct softening of the pressure in the star interior induced by the hyperons.  

Shown in Fig.~\ref{fld3} is the tidal deformability of neutron
stars as a function of neutron star mass $M$. All curves with and
without hyperon fractions pass through the experimental constraint extracted for the 1.4$M_{\odot }$ neutron star which is $190_{-120}^{+390}$~\cite{Abb18}.
With the inclusion of hyperons, the curves move downwards to the
centroid point or further to the lower experimental bound. More
displacement of the curves is observed with larger hyperon
fractions included. For quantitative clarity,
we tabulate the tidal deformability of the 1.4$M_{\odot }$ star in Table~\ref{tab2}. One can see, for instance, that the case with $f_{\sigma\Lambda }=0.8$ and $\alpha =-0.05$ in the third column where the inclusion of 3\% hyperons reduces the tidal deformability from 394.6 to 310.5 by 21.2\%, while an inclusion of 10.7\% hyperons (in the last column with $\alpha =0.05$) can reduce the tidal deformability by 42.4\%. Such a reduction
of tidal deformability can be larger for moderately heavy neutron
stars within a rough mass range of 1.4$M_{\odot }<M<1.6$$M_{\odot
}$, as can be observed in Fig.~\ref{fld3}. Here, we further mention the role of the parameter $\alpha$.
As shown in Table~\ref{tab2} and Figs.~\ref{fmsr} and \ref{fld3}, the roles of the parameters $\alpha$ and $f_{\sigma Y}$ are comparable in adjusting various quantities. We note, however, the small parameter $\alpha$ is more sensitive to adjust the hyperon fractions, see Table~\ref{tab2}, and can efficiently readjust the effect induced by the parameter  $f_{\sigma Y}$, see Fig.~\ref{fmsr}.  

\begin{table}[tbh]
\caption{The tidal deformability $\Lambda _{g}$ of the
1.4$M_{\odot }$ star, together with the hyperon number fraction
(in percentage) in brackets. Without hyperons, $\Lambda _{g}$ is
394.6 and 229.5 with the SLC and SLCd, respectively.} \label{tab2}
\begin{center}
\begin{tabular}{ccccc}
\hline\hline
Model & $f_{\sigma\Lambda}$ & $\alpha=-0.05$ & $\alpha=0$ & $\alpha=0.05$ \\
\cline{2-5}
SLC & 0.8 & 310.6 (3.06) & 255.2 (5.98) & 227.4 (10.7) \\
& 0.9 & 358.8 (1.07) & 319.5 (2.65) & 260.9 (5.33) \\ \cline{2-5}
SLCd & 0.8 & 215.0 (1.90) & 193.1 (4.07) & 163.5 (7.72) \\
& 0.9 & 226.7 (0.47) & 216.4 (1.55) & 195.1 (3.61) \\ \hline\hline
\end{tabular}%
\end{center}
\end{table}
The $\Lambda _{g}-M$ relation, also see Eq.(\ref{eqldg}), is advantageous for observing the effect of the radius difference from various models at the given star mass, since in this case the various models may have close Love numbers in a large mass range.  
This is especially true for models SLC and SLCd that only differ in the density dependence of the symmetry energy and consequently the
radius of normal neutron stars with the simplest compositions
(e+$\mu $+n+p). Without hyperons, the difference
between the $\Lambda _{g}$'s from the SLC and SLCd in the mass region $M>1.3M_{\odot }$ can be specified by a $R^{5}$ dependence as given by Eq.(\ref{eqldg}), as we see from Fig.\ref{fk2} that the corresponding
Love number nearly overlaps for two models. As the hyperons are included, the $\Lambda _{g}$ difference from two models reduces clearly with increasing the hyperon fraction in neutron stars.  This reduction is dominated by the correspondingly reduced radius difference from two models in the presence of hyperons, see Fig.~\ref{fmsr} and Eq.(\ref{eqldg}).   Secondly, the Love
number undergoes a nonlinear decrease with the increase of the
hyperon fraction, see Fig.~\ref{fk2}, and this further reduces the
difference of $\Lambda _{g}$ dominated by the $R^{5}$ dependence.

As a result of the reduced $\Lambda _{g}$ difference in the models
SLC and SLCd that just differ in the density dependence of the
symmetry energy, the symmetry energy constraint extracted from the
gravitational wave data will be quite different with and without
the consideration of hyperon compositions. This is not surprising
because the symmetry energy is determined by the in-medium strong
interactions and the inclusion of hyperons changes the in-medium effect. As shown in Fig.~\ref{fld3}, the $\Lambda _{g}$ with the SLC and SLCd come closer with the moderate
lowering of the hyperon onset density accompanied by an rise of
hyperon fraction. In an extreme case where the $\Lambda _{g}$
overlaps, the sensitivity to the symmetry energy vanishes, which
means that the experimental bounds of the tidal deformability can't be employed to constrain the symmetry energy at all. Generally, the
inclusion of more hyperons can smear out the constraints on the
symmetry energy. The lowest onset density in the present model
parametrizations is 2.57$\rho _{0}$. While other models may have
the onset density as low as 2$\rho _{0}$ or smaller, even much
severer smearing-out can be expected to occur for the symmetry
energy constraint. On the other hand, were the experimental bounds
extracted for heavier stars whose high density content occupies
larger fraction, it is advantageous to obtain the constraints for
hyperon component in stars, as implied from the results in
Fig.~\ref{fld3}. For the constraint on the symmetry energy, it
favors the experimental bounds of a lower mass star. In order to
decouple the effects from the symmetry energy and hyperon
component, measurements of multi-fiducial mass points seem to be
necessary in future gravitational wave experiments. At last, we
clarify that other non-nucleonic degrees of freedom, in addition
to the hyperon component, may also affect the extraction of the
density dependence of the symmetry energy. For instance, the
inclusion of appropriate dark matter candidates can affect the
relation between the mass-radius trajectory of neutron stars and
the symmetry energy~\cite{Xia14}.

\section{Summary}
\label{summary} In a multi-messenger era, the gravitational wave
from neutron star mergers provides a novel probe to the neutron
star interior and the relevant nuclear EOS. In this work, we
utilize the RMF models with the density-dependent parametrizations
to study the neutron star tidal deformability with and without the
inclusion of the hyperon fraction.
With a small parameter $\alpha$, the in-medium vector potential for hyperons is adjusted to affect the onset densities and fractions of hyperons sensitively. The decreased (increased) onset density can result in a clear enhancement (suppression) of the hyperon fraction. We have found that the shift of the hyperon fraction with various onset densities (within $0.3\rho_0$) can sensitively
affect the star tidal deformability. The present results indicate
that the gravitational wave can signal the interior structure and
composition of neutron stars. On the other hand, the complication
also arises since the constraint on the nuclear symmetry energy,
extracted from the gravitational wave signals, would depend on the
scenarios of neutron star composition. According to the present
results with and without inclusion of hyperons, such a dependence
is not negligible.    In particular, the difference in symmetry energies signaled by the star tidal deformability may be largely smeared out by the inclusion of hyperons, though the concrete  result relies on the values of the free parameters ($\af$ and $f_{\sg Y}$)  and is strongly model-dependent.  To distinguish the effect of the symmetry
energy and the hyperon component,  the measurement of
multi-fiducial mass points is thus necessary in future gravitational
wave experiments.

\section*{Acknowledgements}

One of the authors (WZJ) thanks Dr. F. J. Fattoyev for providing us
his code which is very helpful to crosscheck our calculation of
the tidal deformability. The work was supported in part by the
National Natural Science Foundation of China under Grant No.
11775049.

\end{document}